\pdfoutput=1

\documentclass[12pt]{article}
\usepackage{amsmath}
\usepackage{graphicx}
\usepackage[font=small,labelfont=bf]{caption}
\usepackage{hyperref}

\DeclareCaptionFont{tiny}{\tiny}

\newcommand\pubnumber{DPF2015-278}
\newcommand\pubdate{\today}

\def\smu{Department of Physics\\
Southern Methodist University, Dallas, Texas 75275, USA}
\def\support{\footnote{Work supported by the Southern Methodist University Department of Physics.}}

\def\Title#1{\begin{center} {\Large #1 } \end{center}}
\def\Author#1{\begin{center}{ \sc #1} \end{center}}
\def\Address#1{\begin{center}{ \it #1} \end{center}}

\newcommand\pubblock{\rightline{\begin{tabular}{l} \pubnumber\\
         \pubdate  \end{tabular}}}
\newenvironment{Abstract}{\begin{quotation}  }{\end{quotation}}
\newenvironment{Presented}{\begin{quotation} \begin{center} 
             PRESENTED AT\end{center}\bigskip 
      \begin{center}\begin{large}}{\end{large}\end{center} \end{quotation}}
\def\Acknowledgments{\bigskip  \bigskip \begin{center} \begin{large}
             \bf ACKNOWLEDGMENTS \end{large}\end{center}}

\def\beq{\begin{equation}}
\def\eeq#1{\label{#1}\end{equation}}
\def\eeqn{\end{equation}}

\def\beqa{\begin{eqnarray}}
\def\eeqa#1{\label{#1}\end{eqnarray}}
\def\eeqan{\end{eqnarray}}

\let\bar=\overbar

\def\Dslash{\not{\hbox{\kern-4pt $D$}}}
\def\dslash{\not{\hbox{\kern-2pt $\del$}}}

\def\msb{{\bar{\ssstyle M \kern -1pt S}}}

\begin{document}
\begin{titlepage}
\pubblock

\vfill
\Title{ManeParse:Mathematica\textsuperscript{\textregistered} Toolbox for PDF Uncertainties and Application to New Physics Searches}
\vfill
\Author{ Eric Godat\support}
\Address{\smu}
\vfill
\begin{Abstract}
As the LHC begins Run 2 at an even higher energy, one of the top priorities will be to search for new particles (possibly from SUSY) at the highest energy scales. In addition to direct production of new particles, they can mix with Standard Model (SM) particles to yield discrepancies from the usual predictions. To distinguish this new physics from old uncertainties, we need tools to precisely quantify the uncertainties of the SM predictions. Here we present a versatile set of utility functions built with Mathematica that is capable of working with a variety of PDF formats including the recent LHAPDF6 format.

This software can perform PDF calculations within the Mathematica framework and compare results from different PDF collaborations. The package includes both the central PDF value as well as the full error sets needed for PDF uncertainty analysis; a variety of sample error definitions are implemented. We demonstrate this package for the case of a new heavy scalar particle production at the LHC.
\end{Abstract}
\vfill
\begin{Presented}
DPF 2015\\
The Meeting of the American Physical Society\\
Division of Particles and Fields\\
Ann Arbor, Michigan, August 4--8, 2015\\
\end{Presented}
\vfill
\end{titlepage}
\def\thefootnote{\fnsymbol{footnote}}
\setcounter{footnote}{0}

\section{Introduction}

Parton Distribution Functions (PDFs) are essential tools for relating theoretical predictions to experimental data.  PDFs represent parameterized fits to data that describe the momentum fraction carried by the constituent components of hadrons.  Different collaborations have different parameterizations and fit different data sets resulting in variations between PDF sets.  Each PDF grid is broken down by Bjorken $x$, hard scattering energy $Q$ and parton flavor.

\section{Purpose}

The purpose of the ManeParse package is to provide a lightweight PDF reader for multiple collaborations' PDF formats, specifically LHAPDF6~\cite{Buckley:2014ana} and CTEQ PDS~\cite{Gao:2013xoa} formats.  The package utilizes a custom, 4-point Lagrange Interpolation routine that is fast, reliable and transparent.  This interpolation allows the discrete grid points to be treated as a continuous function.  The package was built in the user-friendly environment provided by Mathematica, which allows the user access to powerful built-in plotting and calculation functions.  Additionally, ManeParse includes multiple error propagation techniques, both Hessian and Monte Carlo, and the ability to calculate observables such as cross section and luminosity.  The ManeParse package consists of a set of four packages, two parsing routines for each format, an error package, and a calculation package that houses the interpolation routine and several user functions.  To operate, the user provides the parser with data files, these then are read into memory as a three dimensional grid.  The calculation package takes this grid and can interpolate any values the user specifies as long as the values fall within the range of the grid.  This function can be treated like any other function in Mathematica at this point and can be plotted or used simply to return a value.

\section{Testing}
ManeParse operates at speeds comparable to the proprietary LHAPDF6 and CTEQ codes, built in C++ and FORTRAN respectively~\cite{Buckley:2014ana}~\cite{Gao:2013xoa}. Both the PDS and LHA parsing routines are able to read in full PDF families in seconds, with individual files being read in a few hundredths of a second. After info and header files are read into memory, they can be accessed quickly to aid in the parsing of data files without needing to be reread, thus speeding up the parsing of full PDF families. The interpolation routine can be called approximately two thousand times per second.  This is necessary for being able to do numerical integrations and plotting, as these rely on multiple calls to the interpolation routine.\\

\noindent To test the validity of the interpolation, we utilized the Sum Rules for PDFs.  These rules state that the sum of momentum fractions for all partons must add to one to account for the total momentum of the hadron. Since these rules take into account each parton over the entire range of momentum fractions, $x$, for a given energy, $Q$, any errors in parsing or interpolating would cause fluctuations in the sum.

\begin{figure}[htb]
\centering
\includegraphics[width=.5\textwidth]{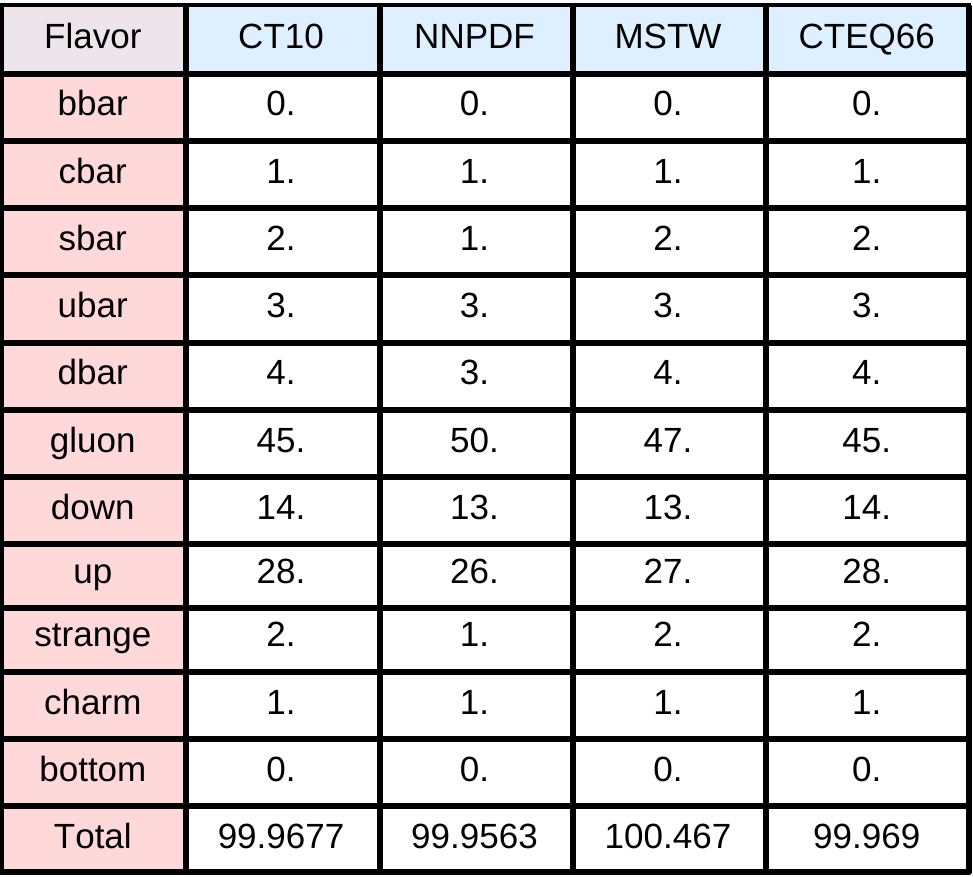}
\captionsetup{font=tiny}
\caption{Sum Rules provide validation for interpolation routine shown here at $Q$ = 2 GeV.}
\label{fig:SumRule_2Gev.pdf}
\end{figure}

\begin{figure}[htb]
\centering
\includegraphics[width=.75\textwidth]{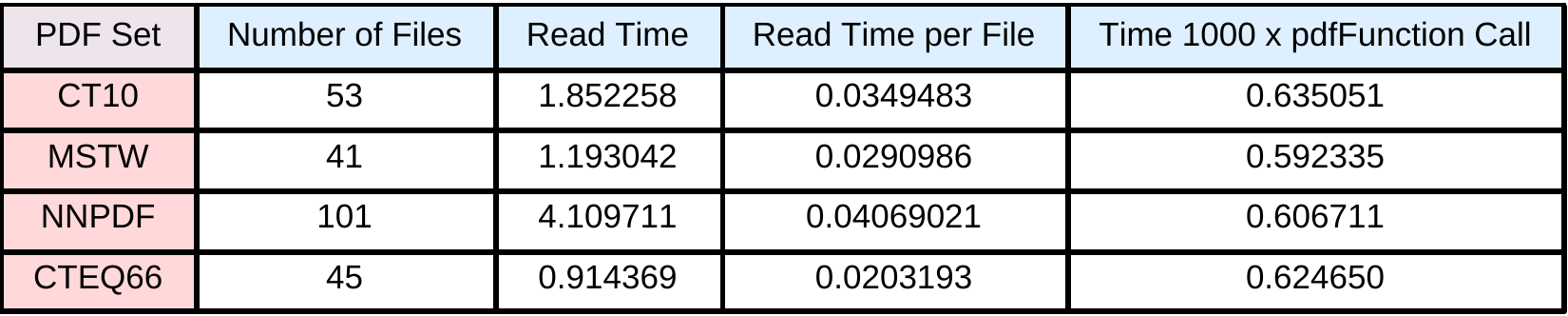}
\captionsetup{font=tiny}
\caption{Timing to read and interpolate for multiple collaborations.}
\label{fig:timingManeParse.pdf}
\end{figure}

\section{Examples}
Here we provide examples of figures generated using ManeParse and built-in Mathematica plotting functions. These plots display the versatility of the Mathematica plotting functions and how easily ManeParse interacts with them. All the examples shown below are a representations of the interpolation routine and/or error functions acting on a PDF set or family of sets.

\begin{figure}[htb]
\centering
\begin{minipage}{.5\textwidth}
  \centering
  \includegraphics[width=.9\linewidth]{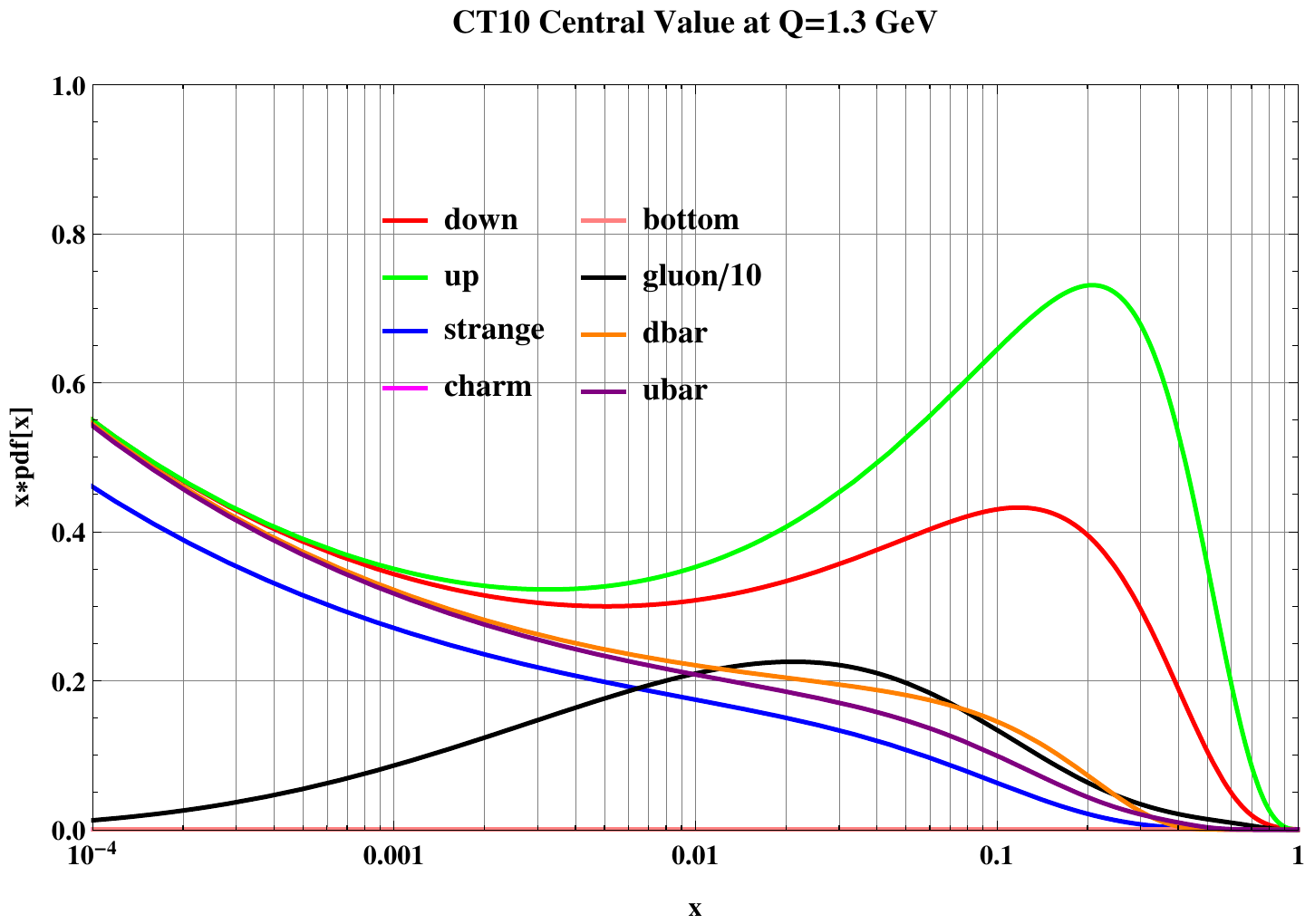}
  \captionsetup{font=tiny}
  \captionsetup{width=.9\textwidth}
  \caption{Plot produced using ManeParse representing multiple flavors over $x$ range.}
  \label{fig:centralvalueCT10_2.pdf}
\end{minipage}%
%
\begin{minipage}{.5\textwidth}
  \centering
  \includegraphics[width=.9\linewidth]{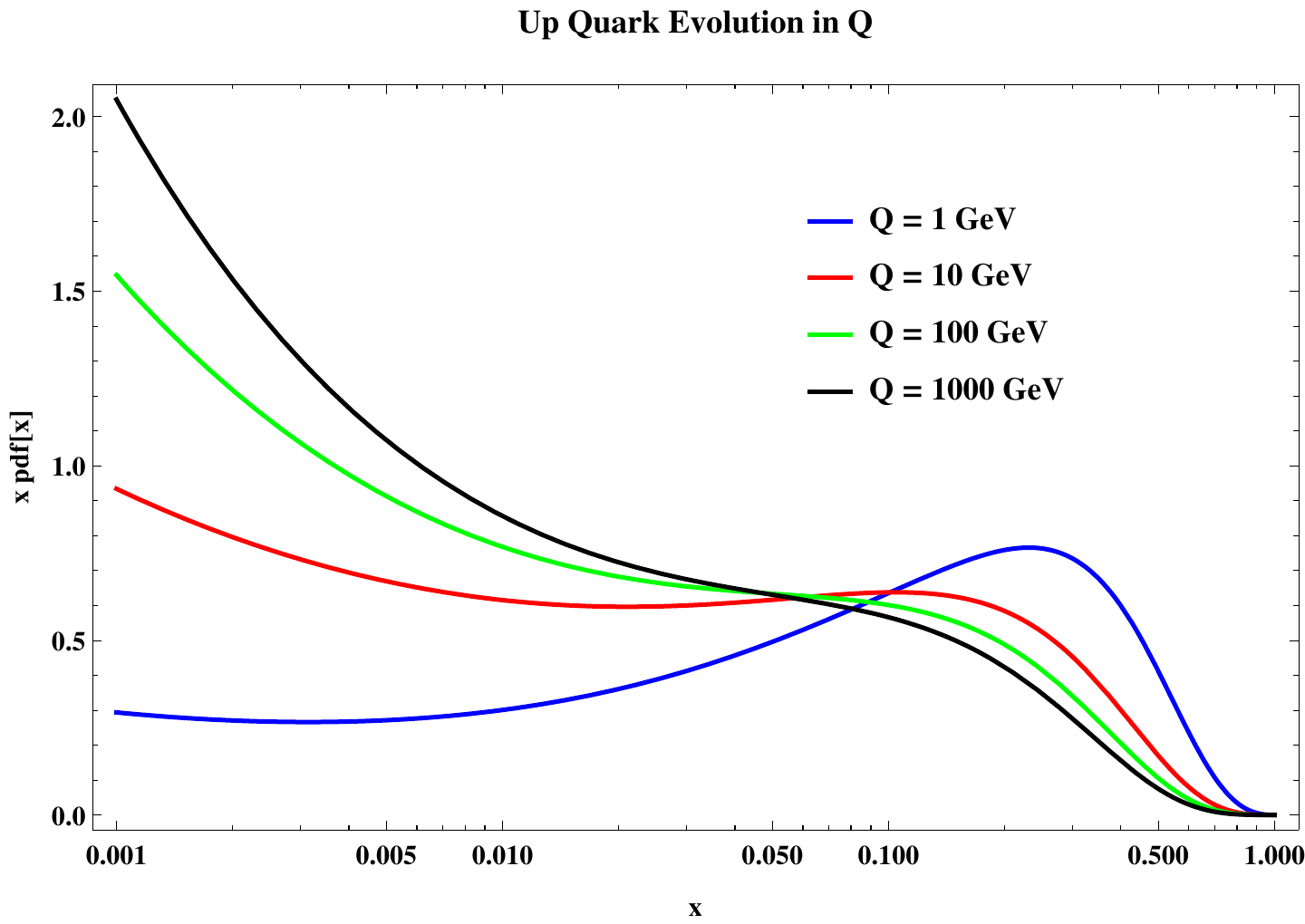}
  \captionsetup{font=tiny}
  \captionsetup{width=.9\textwidth}
  \caption{Plot showing up PDF for increasing $Q$.}
  \label{fig:Up_evo_plot_2.pdf}
\end{minipage}
\end{figure}
%
\begin{figure}[htb]
\centering
\begin{minipage}{.5\textwidth}
  \centering
  \includegraphics[width=.9\linewidth]{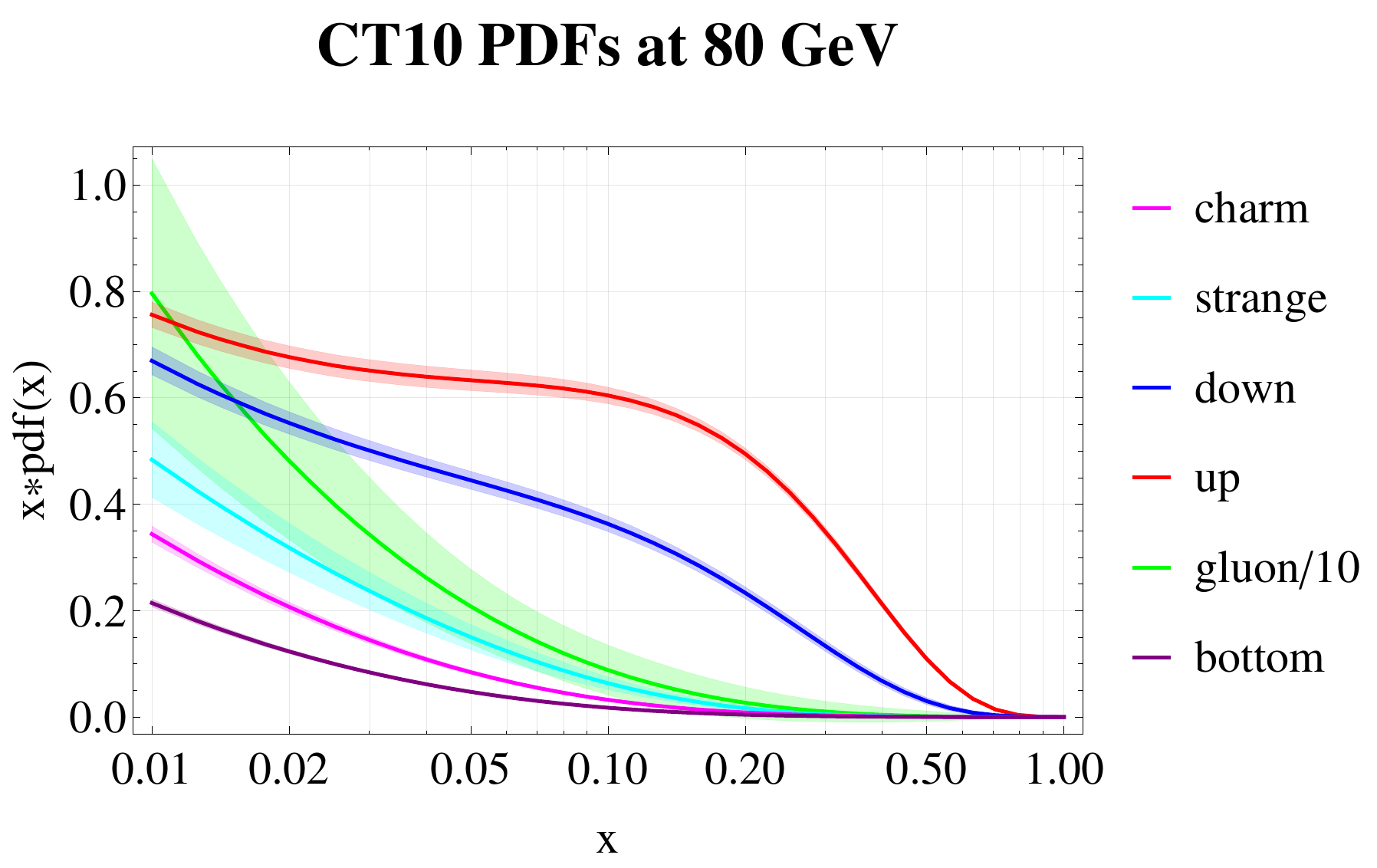}
  \captionsetup{font=tiny}
  \captionsetup{width=.9\textwidth}
  \caption{Plot including error bands for multiple flavors over $x$ range.}
  \label{fig:ct10Plots.pdf}
  \end{minipage}%
%
\begin{minipage}{.5\textwidth}
  \centering
  \includegraphics[width=.9\linewidth]{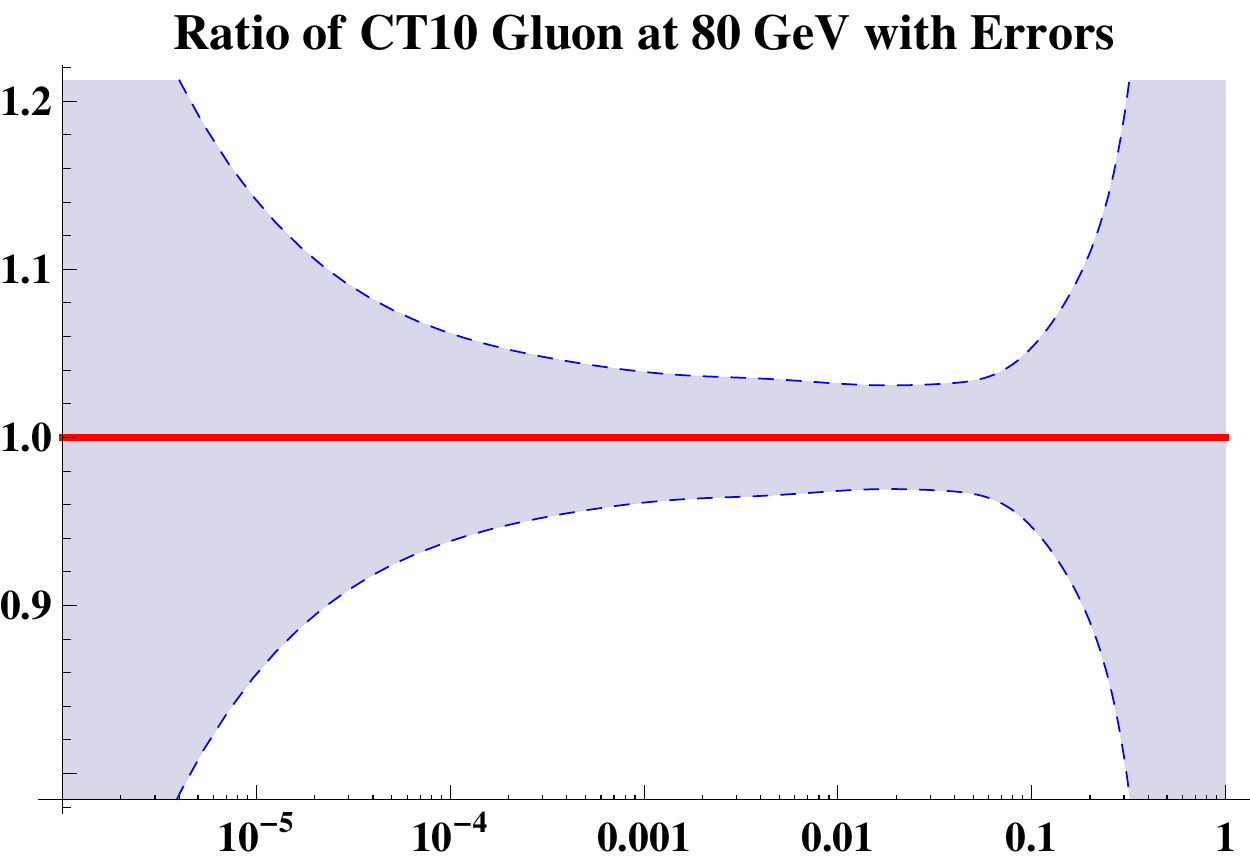}
  \captionsetup{font=tiny}
  \captionsetup{width=.9\textwidth}
  \caption{Ratio plot of gluon errors bands to gluon central value PDF.}
  \label{fig:ratioerrorPlot.pdf}
\end{minipage}
\end{figure}
%
\begin{figure}[htb]
\centering
\begin{minipage}{.5\textwidth}
  \centering
  \includegraphics[width=.9\linewidth]{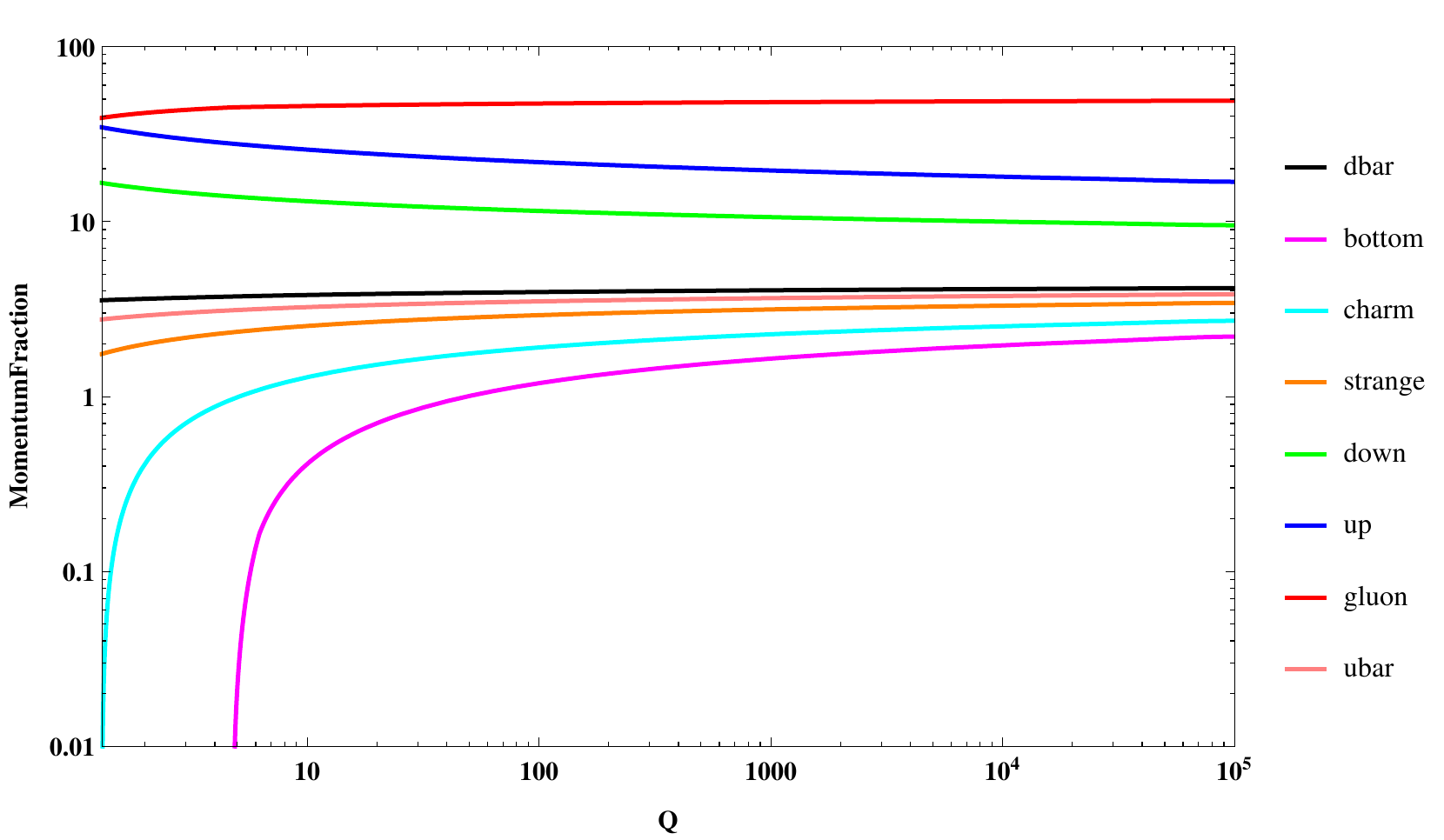}
  \captionsetup{font=tiny}
  \captionsetup{width=.9\textwidth}
  \caption{Plot showing momentum fractions for increasing $Q$. This plot reflects what is calculated in the Sum Rules.}
  \label{fig:momfracplot.pdf}
  \end{minipage}%
%
\begin{minipage}{.5\textwidth}
\centering
\includegraphics[width=.9\linewidth]{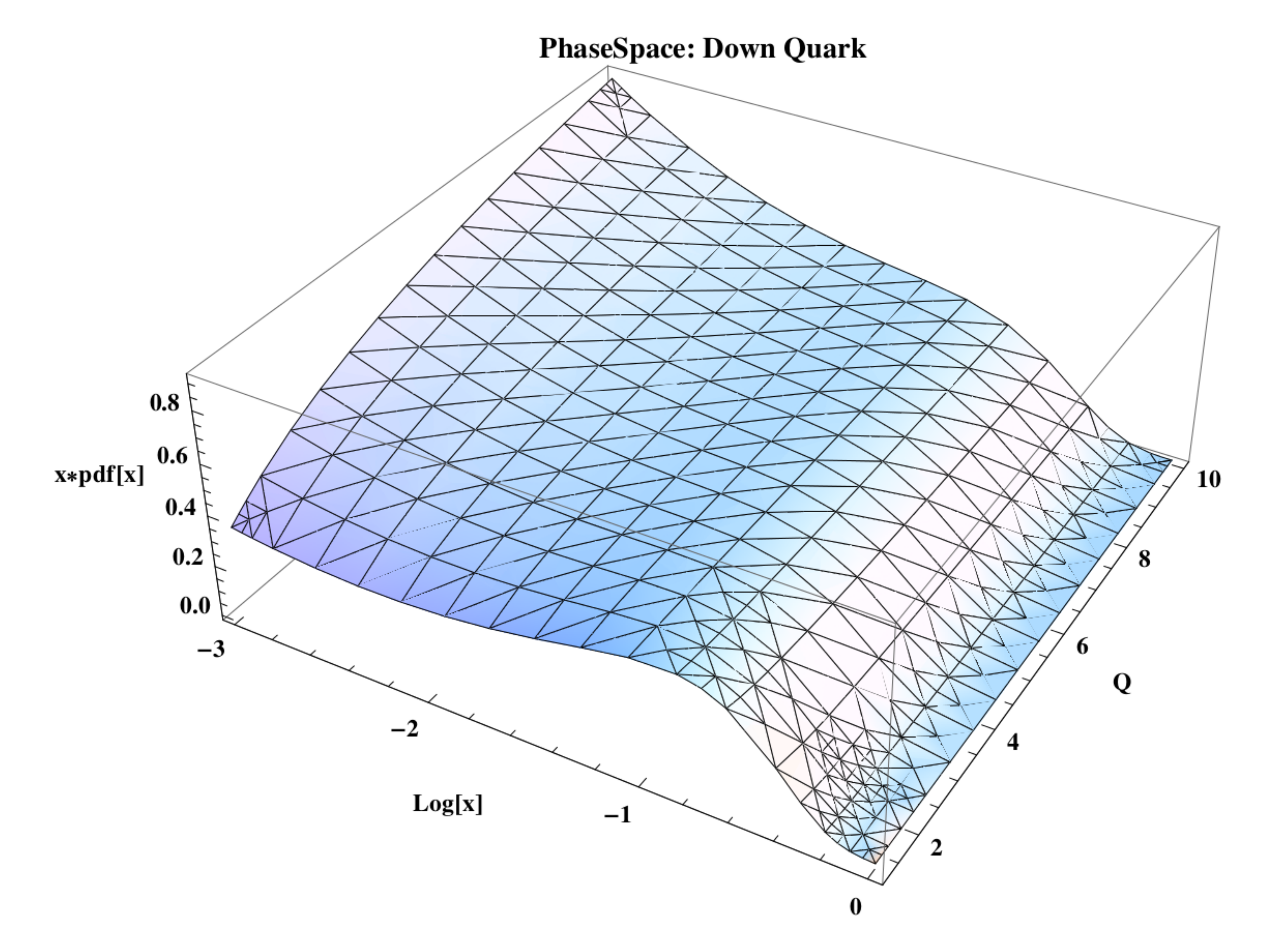}
\captionsetup{font=tiny}
\caption{Visualization of Phase Space are possible with Mathematica 3D Plotting.}
\label{fig:PhaseSpaceDownQuark.pdf}
\end{minipage}
\end{figure}
\clearpage
\section{Applications to New Physics Searches}

ManeParse has several applications. First, we discuss an examination of intrinsic bottom quark contribution within the proton at the LHC by utilizing the ability to calculate luminosities, and second, we provide a brief description of the ability to extrapolate at low values of $x$ for examining ultrahigh energy neutrinos.\\

\noindent Luminosities provide the link between experimental and theoretical cross section measurements,  
\[
\sigma \left(\hat{s}\right)~=~\sum _{ij}\int\limits _{\tau}^1\frac{d\mathcal{L}_{ij}}{d\tau }~\hat{\sigma }_{ij}\left(\hat{s}\right)d\tau \quad.
\]
Integrated luminosity can be written as a convolution of PDFs and thus can be easily calculated with ManeParse.  Here the speed of the interpolation routine makes the numerical integration practical and efficient,
\[
\frac{d\mathcal{L}_{ij}}{d\tau }(\tau ,\mu )=\frac{1}{\delta _{ij}+1}\int\limits_{\tau}^1 \frac{1}{x}\left[f_i(x,\mu )f_j\left(\frac{\tau }{x},\mu \right)+f_j(x,\mu )f_i\left(\frac{\tau }{x},\mu \right)\right] \, dx \quad.
\]
Using the assumptions made by \cite{Lyonnet:2015sda}, we are able to use ManeParse to reproduce their results and compare them to other collaborations' PDF sets as well.\\

\noindent Additionally, we were able to use a custom-made intrinsic bottom quark PDF~\cite{Lyonnet:2015sda}, utilizing the decoupling of the extrinsic and intrinsic PDFs due to the DGLAP evolution equations, to directly add the intrinsic component to the existing extrinsic bottom PDF in various amounts to determine the efficacy of separating this effect from the error.\\

\begin{figure}[htb]
\centering
\begin{minipage}{.5\textwidth}
  \centering
  \includegraphics[width=.9\linewidth]{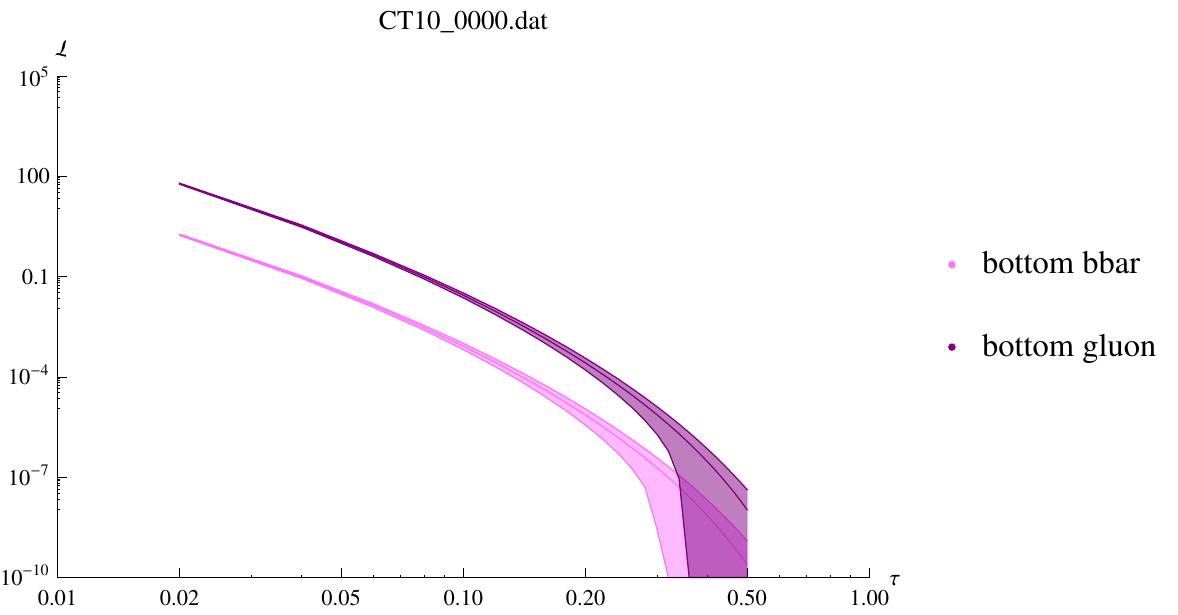}
  \captionsetup{font=tiny}
  \captionsetup{width=.9\textwidth}
  \caption{Luminosities with error bars.}
  \label{fig:luminositywerrors.pdf}
  \end{minipage}%
%
\begin{minipage}{.5\textwidth}
\centering
\includegraphics[width=.9\linewidth]{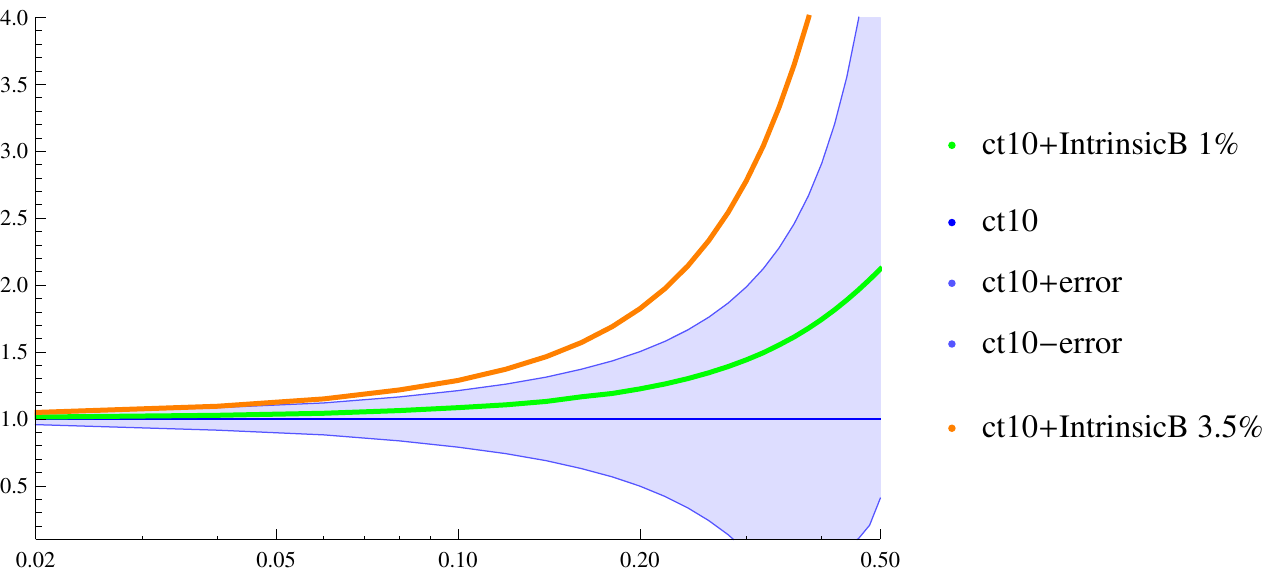}
\captionsetup{font=tiny}
\caption{Ratio Plot of CT10 Luminosities with added Intrinsic Bottom Quark component.}
\label{fig:intrinsicCT10.pdf}
\end{minipage}
\end{figure}\clearpage

\noindent For the case of ultrahigh energy neutrinos, we require very low $x$, much lower than there exists data. ManeParse allows the user to extrapolate to these low $x$ values and tune the powers of $x$ in this region to better fit the particular model~\cite{Gandhi:1998ri}.\\

\begin{figure}[htb]
\centering
\includegraphics[height=1.5in]{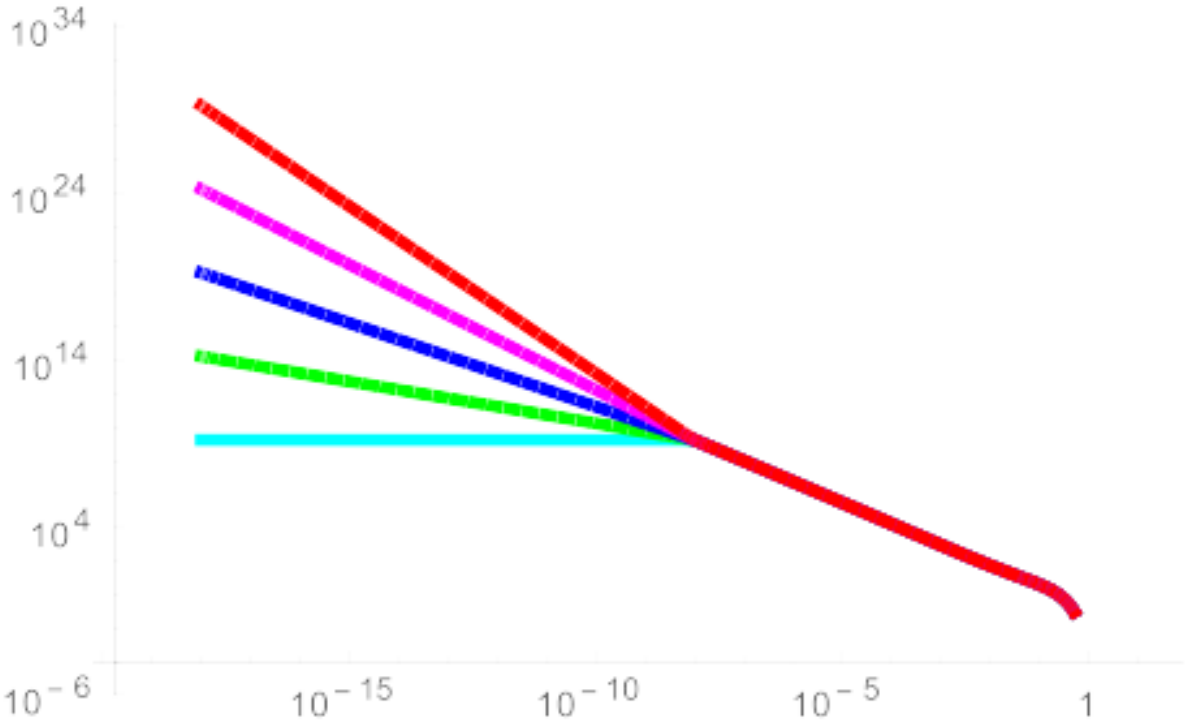}
\captionsetup{font=tiny}
\caption{Extrapolation of the PDF into very low $x$ region at various powers of $x$ where PDF value is a function of $x$}
\label{fig:ultrahigh_neutrino.pdf}
\end{figure}

\noindent ManeParse is ready for release and is available for download. It comes with a complete Demo notebook that provides the user sample plotting routines and calculations.

\noindent To Download please visit: \url{ncteq.hepforge.org/code/pdf.html}\\

\Acknowledgments
Thank you to Fred Olness, Florian Lyonnet, Aleksander Kusina and Ben Clark with whom this work was completed.

\end{document}